\documentclass[aps,prl,twocolumn,groupedaddress,showpacs]{revtex4}

\usepackage{longtable}
\usepackage{graphicx}
\usepackage{dcolumn}
\usepackage{bm}
\begin{document}

\title{Transition from a mixed to a pure \emph{d}-wave symmetry in superconducting optimally doped YBa$_2$Cu$_3$O$_{7-x}$
thin films under applied fields}

\author{G. Elhalel}
\author{R. Beck}
\author{G. Leibovitch}
\author{G. Deutscher}
\email {guyde@post.tau.ac.il} \affiliation{School of Physics and
astronomy, Raymond and Beverly Sackler Faculty of Exact Sciences,
Tel-Aviv University, Tel Aviv, 69978, Israel}

\date{\today}

\begin{abstract}
We have probed the Landau levels of nodal quasi-particles by
tunneling along a nodal direction of (110) oriented
YBa$_2$Cu$_3$O$_{7-x}$ thin films with a magnetic field applied
perpendicular to the $CuO_2$ planes, and parallel to the film's
surface. In optimally doped films and at low temperature, finite
energy nodal states are clearly observed in films thinner than the
London penetration depth. Above a well defined temperature the
order parameter reverts to a pure \emph{d}-wave symmetry.
\end{abstract}


\maketitle


The pairing interaction responsible for superconductivity in the
high temperature superconductors is still under debate. There is
much experimental evidence showing that the order parameter has a
dominant $d_{x^2-y^2}$ symmetry \cite{Hardy,Tsuei}, but important
questions such as the nature of quasi-particle states in the
under-doped region and the existence of a minority component of
the order parameter are still under debate \cite{Deutscher}.
Studying the density of states when a magnetic field is applied
along the c-axis can lend important information on both of these
questions.
\par

Gor'kov and Schrieffer \cite{gorkov} have remarked that in a
$d$-wave superconductor nodal quasi-particles undergo in their
orbital motion a series of Andreev - Saint-James reflections  from
nearby lobes of the order parameter, resulting in finite energy
states. This states, as noted by Anderson \cite{Anderson}, carry a
current around the Fermi surface. Each time an electron, say, is
reflected as a hole, the missing pair reappears as a Cooper pair
in the condensate. Such currents can be described by an $id_{xy}$
component which, as shown by Laughlin \cite{Laughlin}, lowers the
free energy of the superconducting state under the applied field
because of the moment they produce. Minimization of the free
energy with respect to the amplitude, $\delta_{xy}$, of the
minority component,  lends at $T=0$ the law: $\delta_{xy}$ =
$aH^{1/2}$ with $a=\hbar\upsilon\sqrt{\frac{2e}{\hbar c}}$. Here
$\upsilon=\sqrt{\upsilon_1 \upsilon_2 }$, where $\upsilon_1$ is
the Fermi velocity,
$v_2=\frac{1}{\hbar}\frac{\partial\Delta}{\partial k}$ at the node
direction, and $\Delta$ is the main $d_{x^2-y^2}$ superconducting
gap. A weak first order phase transition to a pure \emph{d}-wave
symmetry was predicted to occur at a temperature $k_BT_{CF} =
b\delta_{xy}(H)$, where $b$ is a universal constant equal to 0.52.
It was however pointed out that in the mixed state Meissner
currents would cause a large Doppler shift of the nodal states
energy, making their observation difficult if not impossible
\cite{Franz}.

\par
In this letter we show how this difficulty can be resolved by
taking measurements on films thinner than the London penetration
depth, and conclude that a field induced minority component exists
below a well defined temperature $T_{CF}$. Our results are in
semi-quantitative agreement with Laughlin's theory \cite{Laughlin}
with proportionality constant, $b=0.34$, smaller than the
predicted one.

\par
One of the most powerful tools for probing the order-parameter is
tunneling spectroscopy. For conventional superconductors, the
tunneling spectrum is directly proportional to the
superconductor's density of state and exhibits two peaks at biases
corresponding to $\pm\Delta$, where $\Delta$ is the
superconducting gap \cite{giaever}. For a $d$-wave order-parameter
symmetry, zero energy surface states are formed when
quasi-particles undergo successive Andreev - Saint-James
reflections from order parameters whose phase differ by $\pi$, as
occurs upon reflection at a surface perpendicular to a nodal
direction. Tunneling along a node direction, \emph{i.e.} the (110)
direction, shows a zero bias peak in the tunneling spectrum
\cite{Tanaka} due to these states \cite{Hu2}. An additional
minority $id_{xy}$ component would show up as peaks of the
junction's conductance at biases equal to $\delta_{xy}$, replacing
the zero bias peak \cite{Tanaka}.

\par
According to theoretical predictions
\cite{gorkov,Anderson,Laughlin}, tunneling spectral peaks should
thus appear when the field is applied perpendicular to the CuO$_2$
planes in films having a nodal orientation. The presence of such
peaks was noted previously, but received a different interpretation
\cite{Lesueur:1992,Covington,FSR}, which is that Meissner screening
currents will Doppler shift the zero energy surface states. It was
shown \cite{Krupke} that the split of the zero bias conductance peak
is always larger in increasing fields than in decreasing ones, a
behavior that can be understood on the basis of the hysteretic
behavior of screening currents. These currents are of two kinds:
Meissner currents on the scale of the London penetration depth, and
Bean currents due to vortex pinning. Meissner currents are large in
increasing fields, due to the Bean Livingston barrier that retards
the penetration of vortices, and small in decreasing fields because
there is no barrier against vortex exit. As for Bean currents, they
reverse sign upon field reversal, and are typically weak compared to
Meissner currents in increasing fields.  In experiments performed on
YBa$_2$Cu$_3$O$_{7-x}$ thick films having the (110) orientation,
Beck \emph{et al.} \cite{Beck:2004} have reported on finite bias
conductance peaks that follow the $H^{1/2}$ law in decreasing fields
where screening currents are weak, and attributed them to a field
induced $id_{xy}$ component. This conclusion has however remained
controversial as the intensity of screening currents is not known
with certainty.
\par
In an effort to establish more firmly the origin of the field
induced finite bias conductance peaks, we have made a series of
tunneling measurements as a function of field and temperature on
films thinner than the London penetration depth. Both the Bean
Livingston and the Bean currents must then be quite weak, since
screening currents generated on opposite faces of the film cancel
each other. If the origin of the finite bias peaks seen in thick
films is a Doppler shift of zero energy surface bound states, they
should not occur at all in thin films, either in increasing or in
decreasing fields. If, on the other hand, they are basically due to
field induced finite energy nodal states, they should persist in the
thin films with no field hysteresis, and follow the $H^{1/2}$. As we
now describe in detail, this is indeed what we have found.
\par
\begin{figure}
\includegraphics[width=0.7\hsize]{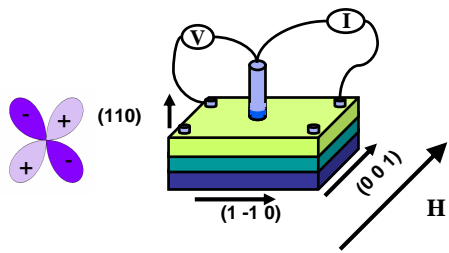}
\caption{\label{figschema}A scheme of the samples' configuration.
The samples' surface is in the (110) direction, and one of the
surface edges is parallel to the c-axis. The magnetic field is
applied parallel to the surface and to the c-axis. }
\end{figure}
\par
We fabricated (110) oriented YBa$_2$Cu$_3$O$_{7-x}$ thin films by
DC off-axis sputtering on (110) SiTrO$_3$ substrates with one of
the surface edges parallel to the c-axis. A
PrBa$_2$Cu$_3$O$_{7-x}$ buffer layer was first deposited by
off-axis RF sputtering in order to reduce the growth of (103)
oriented grains \cite{Poelders}. The thickness of the samples is
less than 500\AA. X-ray diffraction reveals peaks corresponding to
the (110) orientation, the $T_C$ of the samples ranged from 84 K
to 89 K down-set. The films' surface is relatively smooth, having
an average surface roughness of few tens of angstroms according to
atomic force microscope measurements. We verified the in-plane
orientation by comparing the normal state resistance along the two
different directions and found the expected anisotropy
\cite{Poelders}.
\par

\begin{figure}
\includegraphics[width=0.9\hsize]{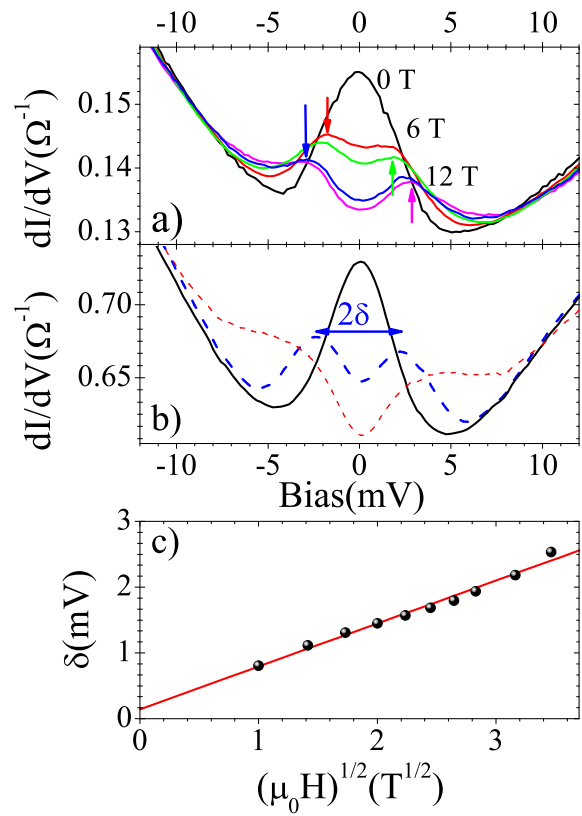}
\caption{\label{fig2}(color online) (a) The conductance spectrum of
a thin sample obtained when increasing and decreasing the magnetic
field. Full black line: H=0 T. Arrows facing up(down) represent
increasing(decreasing) field. (b) The conductance spectrum obtained
on a thick sample. Full black line: H=0, thin dashed line: H=4 T
increasing field, thick dashed line: H=4T decreasing field.(c)
Spectral peak value versus the square root of applied field in
decreasing fields measured on a thin sample at 4.2 K. The solid line
is a linear fit with a slope of 0.65 mV/T$^{1/2}$. We note that the
slope found here is slightly smaller than that found for thicker
films (see text).}
\end{figure}
\par
The tunneling junctions are produced by pressing Indium thin pads on
fresh samples, as described elsewhere
\cite{Krupke,DeutscherPhysicaC}. The configuration of the sample and
the field orientation is shown in Fig. \ref{figschema}. The
Indium-Oxide layer, which is created in the area of contact, results
in junction resistances in the range of 1-15 $\Omega$. This process
may reduce the oxygen concentration in the junctions' area,
especially in very thin samples. In order to alleviate this effect,
we grew our films in the presence of a reduced vapor pressure of
water, which results in increased oxygen content in the film
\cite{DaganPRL}. The junctions are stable under thermal cycles, and
the junctions' resistance remains unchanged if kept in a helium gas
environment. This allows us to measure the temperature dependence of
their tunneling characteristic, if necessary by performing
successive cooling runs under various applied fields. The planar
junction technique averages the tunneling conductance at a
macroscopic length scale in contrast to scanning tunneling
microscopy measurements which give sharper peak features \cite{Wei}.
However, its stability over thermal and magnetic fields cycles is of
great advantage for the  present work.
\par
In Fig. \ref{fig2}a we show the differential conductance obtained on
a thin film at 4.2 K, measured at magnetic fields $H$=0, 6 and 12 T,
in increasing and decreasing fields. At zero magnetic field, there
is a clear peak at zero bias, as expected for a (110) oriented film
\cite{Tanaka}. This peak splits into two spectral peaks in the
presence of a finite magnetic field. The conductance spectra in
increasing and decreasing fields are almost identical as opposed to
the high hysteretic behavior of the spectral peaks in thicker
samples as shown in Fig. \ref{fig2}b and in Ref.
\cite{Krupke,DeutscherPhysicaC}. This highly hysteretic behavior has
been ascribed to strong screening currents in increasing fields and
much weaker ones in decreasing fields as discussed above.
\cite{BeckHystersis,Beck:2004}. The negligible hysteresis observed
in our thin samples implies that the screening currents are very low
even in increasing fields and that the two finite bias peaks cannot
be due to a Doppler shift of the zero energy bound states. The
difference between the behavior of thin and thick films is
particularly dramatic in increasing fields, where the spectral peaks
cannot be identified anymore at all in thick samples. This is
because the Doppler shift, due to Meissner currents, widens the
peaks and pushes them into high bias where they are merged with the
background and the main gap structure (Fig. {\ref{fig2}}b).

\par
We define the minority gap peak value, $\delta$, as half the
distance between the positive and negative bias conductance peaks
(see Fig. \ref{fig2}b). In our thin films, $\delta(H)$ follows the
$H^{1/2}$ behavior as shown in Fig. \ref{fig2}c. This is in
agreement with measurements performed earlier on thick films in
decreasing fields \cite{Beck:2004,BeckHystersis} and confirms the
assumption that Meissner currents are weak in that case, but may be
not negligible in view of the smaller slope obtained here.
\par
\begin{figure}
\includegraphics[width=1\hsize]{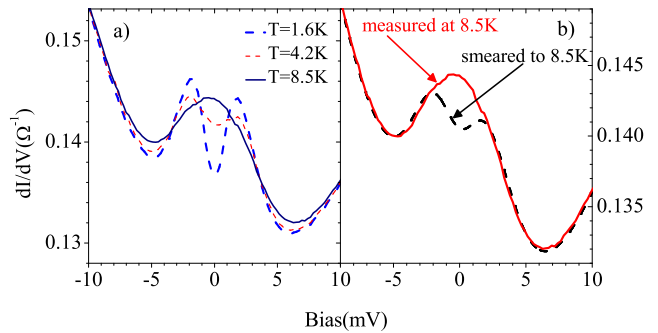}
\caption{\label{fig3}(color online) (a) Conductance measured under
a field of 7 T at various temperatures. (b) Comparison between the
measured data at $T$=8.5 K, $H$=7 T(full line) and the result of
smearing the data measured at $T$=1.6 K, $H$=7 T to $T$=8.5 K
(dashed line).}
\end{figure}

Measuring the tunneling spectrum at various magnetic fields and
temperatures, we find that in a fixed magnetic field, the minority
gap peak value disappears above a well defined temperature. For
example, we show in Fig. \ref{fig3}a the measured conductance in a
field of 7 T at various temperatures. At 1.6 K, there are two
clear peaks at $\pm$ 1.75 mV. As the temperature rises, the
conductance at zero bias increases until the finite bias peaks
completely disappear at 8.5 K. The conductance at high biases is
independent of temperature as expected for tunneling, assuring us
that the junction characteristics remain intact.
\par
\par
However, in order to be able to properly assert if there is a
temperature induced modification of the density of states at low
bias, one must take into account the effect of thermal smearing on
the tunneling spectra \cite{giaever}. Therefore, we have convoluted
the tunneling spectra measured at the lowest temperature (below 1.6
K) with the derivative of the Fermi-Dirac distribution function at
the desired temperature. Previous scanning tunneling measurements
showed that the density of states hardly changes at such low
temperatures \cite{fisher}. Therefore, using our procedure, we have
calculated a thermally smeared curve for higher temperatures and
thus estimated the effect of thermal smearing on the low energy
spectrum. Figure \ref{fig3}b shows the measurement taken at 8.5 K
(red full line) and the result of smearing the 1.6 K measurement to
8.5 K (black dashed line). While the smeared curve still shows two
distinct peaks, the measurement does not, but rather exhibits a peak
at zero bias. It is therefore clear that temperature induces a
change in the density of states. At biases higher than 5 mV, the two
curves are identical.
\par
\begin{figure}
\includegraphics[width=1\hsize]{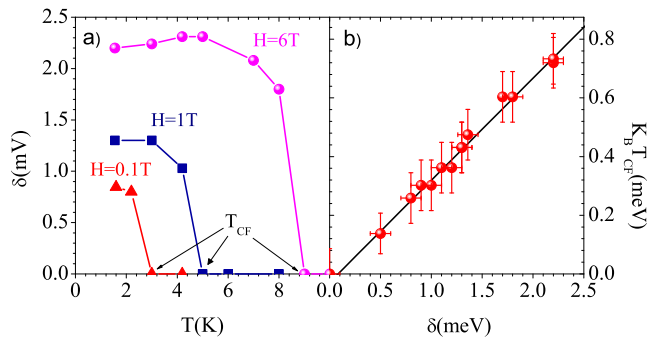}
\caption{\label{fig4}(color online) (a) The spectral peak values
versus temperature obtained from different applied magnetic
fields. (b) The temperature at which the spectral peak disappears
versus the peak value at low temperatures.}
\end{figure}

A representative graph of the minority gap value versus temperature
for various values of the applied field is shown in Fig.
\ref{fig4}a. The minority gap value is approximately constant at low
temperatures and disappears abruptly at a temperature, which we
define as $T_{CF}$. The minor initial rise in the sub gap peaks'
value is an effect of thermal smearing.
\par
The value of $T_{CF}$ increases with the applied field and the
resulting low temperature conductance peak value. We find that they
obey the universal linear relation $k_BT_{CF}=0.34\delta$ as shown
in Fig. \ref{fig4}b. The data points in the graph were taken from
five different samples with different characteristics (one of the
points was measured on a thick sample in decreasing field), which
implies the universality of the proportionality constant. The
measured slope is smaller than Laughlins' prediction
$k_BT_{CF}=0.52\delta$. Laughlins' calculation was performed for
bulk materials, while our measurements were done on very thin films,
which are highly susceptible to surface effects. Laughlin did not
mention the length scale of the currents responsible for the dipole
moments. If the thickness of the film is comparable to the
aforementioned length scale, interaction between the currents
resulting in a partial cancelation of the opposite currents should
be considered. This might contribute to the disagreement between the
theoretical ratio and our experimental result.
\par
A way to distinguish between the Doppler shift effect and the
minority component induced by the applied field, is by their
respective temperature dependence. A first order phase transition
that leads to the abrupt disappearance of the minority order
parameter at high temperatures has been predicted \cite{Laughlin}.
On the other hand, screening currents and the accompanying Doppler
shift should persist at high temperatures since the superfluid
momentum is independent of temperature at temperatures significantly
lower than $T_C$ \cite{Djupmyr}. In addition to the small field
hysteresis in the position of the finite bias peaks, their sharp
disappearance at a well defined temperature rules out that a Doppler
shift of the zero energy surface bound states is at their origin.

\par
The proposed interpretation of the experiments that we have
described requires that the scattering time of nodal quasi-particles
be long compared to the time it takes them to complete a Saint-James
cycle. This appears to be true at optimum doping, but not away from
it as we show in a separate publication \cite{BeckNew}.
\par
In conclusion, we claim that the field induced spectral peaks seen
in nodal tunneling in very thin films result from the formation of a
minority order parameter, and cannot be attributed to a Doppler
shift of the zero energy states due to Meissner screening currents.
The square root dependency of the spectral peak positions in the
magnetic field strength, the absence of field hysteresis and their
abrupt disappearance with temperature are all in favor a field
induced $id_{xy}$ minority imaginary order parameter at low
temperatures and a transition to a pure \emph{d}-wave order
parameter at high temperatures, as predicted by Laughlin. Our
findings are in general agreement with his model, but a comparison
of the way in which the tunneling density of states evolves with
temperature would require a more detailed theory than is now
available. The absence of a minority order parameter inferred from
the heat capacity square root field dependence \cite{Moler} is not
in contradiction with our conclusions, because these measurements
were performed on much thicker crystals where superfluid currents
are strong (and at the origin of the square root dependence) and
predicted to render the minority component unobservable
\cite{Franz}.

\begin{acknowledgments}
This work was supported by the Israel Science Foundation and the
Heinrich-Hertz Minerva Center for High Temperature
Superconductivity. A portion of this work was performed at the
NHMFL, which is supported by NSF Cooperative Agreement No.
DMR-0084173, by the State of Florida, and by the DOE. We thank
Alex Gerber assistance in the high magnetic field measurements.
This manuscript was prepared in part at Stanford University whose
support is gratefully acknowledged by one of us (G.D.).

\end{acknowledgments}

\bibliographystyle{apsrev}

\bibliography{galText}

\end{document}